\documentclass[twocolumn,showpacs,amsmath,amssymb,prl]{revtex4}
\usepackage{graphicx}
\usepackage{dcolumn}
\usepackage{bm}
\begin{document}
\input{epsf}

\title{Spurious Velocities in Dynamically-Cold Systems Due to \\ the
Gravitational Redshifts of their Constituent Stars}

\author{Abraham Loeb}

\affiliation{Astronomy Department, Harvard University, 60 Garden St.,
Cambridge, MA 02138, USA}

\begin{abstract} 

The Doppler effect is commonly used to infer the velocity difference
between stars based on the relative shifts in the rest-frame
wavelengths of their spectral features. In dynamically-cold systems
with a low velocity dispersion, such as wide binaries, loose star
clusters, cold stellar streams or cosmological mini-halos, the scatter
in the gravitational redshift from the surface of the constituent
stars needs to be taken into account as well. Gravitational redshifts
could be important for wide binaries composed of main sequence stars
with separations $\gtrsim 10^{-2}$ pc or in mini-halos with velocity
dispersions $\lesssim 1~{\rm km~s^{-1}}$. Variable redshift could also
lead to a spurious ``detection'' of low-mass planets around a star
with periodic photospheric radius variations.

\end{abstract}

\date{\today}
\maketitle

\paragraph*{Introduction.} According to General Relativity, 
the spectrum of radiation emitted from the surface of a star is
gravitationally redshifted relative to a distant observer.  In the
weak field regime, the spectroscopic velocity shift $v$ is given by
\cite{WeinbergS},
\begin{equation}
v=-{GM\over c R} = - 0.636 \left({M\over M_\odot}\right)
\left({R\over R_\odot}\right)^{-1}~{\rm km~s^{-1}} ,
\label{velocity}
\end{equation}
for a star of mass $M$ and photospheric radius $R$.  The observed
radii of stars show significant scatter for masses $\gtrsim 1 M_\odot$
due to stellar evolution and rotation (see Figures 2 and 17 in
Ref. \cite{Torres}), implying that redshift variations of $\sim
0.5~{\rm km~s^{-1}}$ should be common in stellar systems with stars of
different masses.

\paragraph*{Implications for wide binaries.} The circular velocity of 
a test particle around a star of mass $M$ falls below the velocity
shift $v$ at orbital radii that exceed a critical value,
\begin{equation}
r_{\rm crit} \equiv {c^2 R^2\over GM}= 3.28 \times 10^{16}
\left({R\over R_\odot}\right)^2\left({M\over M_\odot}\right)^{-1}~{\rm
cm} ,
\label{distance}
\end{equation}
Hence, the spectroscopic velocity difference between the stellar
members of a wide binary separated by more than $\sim 10^{-2}$ pc
could be significantly affected by the difference in their intrinsic
gravitational redshifts. Eccentric binaries spend a substantial
fraction of their orbital time near apocenter where their relative
speed is small and the fractional impact of the gravitational redshift
effect is maximized.  At separations larger than $2r_{\rm crit}$, the
binary may be incorrectly declared as unbound if the gravitational
redshift differential is not corrected for.  Very wide binaries had
been studied in the past \cite{Heggie, Bahcall, Retterer, Weinberg,
Quinn}, but the spectroscopic accuracy of their measured radial
velocities \cite{Shaya} was too poor for noticing the stellar redshift
effect. A larger signal of gravitational redshift had already been
detected for compact stars, such as white dwarfs \cite{Greenstein,Barstow,
Falcon,Maoz} and neutron stars \cite{Cottam}.

\paragraph*{False planets.} Long-term periodic variations in the  
photospheric radius of a star, $R(t)$, could lead to a spurious
``detection'' of a low mass planet next to it due to the time
dependence of the gravitational redshift \cite{Cegla}. Such variations
could be induced by the variable suppression of convection due to a
changing number of spots during a magnetic cycle
\cite{Beckers,Kaisig}, or by a change in the number of spots over a
rotation period \cite{Boisse}. Processes like these could modulate in
time the effective radius $R$ of the emission surface, averaged across
the observed face of the star. In principle, the variable
gravitational redshift could dominate over the variable Doppler effect
of the moving photosphere for long periods of variation. The apparent
(false) acceleration of the star would be,
\begin{equation}
\dot{v}={GM\over cR^2}\dot{R} .
\label{dotv}
\end{equation}
If not accounted for, this false acceleration could be incorrectly
associated with a planetary companion at an orbital radius $r$ such
that its orbital period would match the photospheric variability
period. The false planet's mass $m$ would then be chosen to reproduce
the inferred acceleration $\dot{v}$ at the inferred orbital separation
$r$. Assuming a circular planetary orbit and equating $Gm/r^2=\vert
\dot{v}\vert$, one gets
\begin{equation}
m = \left({r\over R}\right)^2\left({\vert \dot{R}\vert\over c}\right) M ,
\label{mass}
\end{equation}
where $\vert \dot{R}\vert =2\pi\Delta R/P$ with $\Delta R$ and $P$ being the
variation amplitude and period.  For example, a periodic change
by merely $\Delta R \approx 1.44\times 10^{-4}R_\odot=100$ km in the
photospheric radius of a Sun-like star with a period of 1 year, would
lead to a false identification of an Earth-mass planet at an orbital
separation of 1 AU. Variable gravitational redshift was likely an
unimportant contaminant in the recent identification of an Earth-mass
planet at an orbital period of 3.24 days around the nearby Sun-like
star $\alpha$ Centauri B \cite{Mayor}.

For searches of low-mass planets in the habitable zone of a Sun-like
star, the false apparent velocity, $\Delta v= - (\Delta R/R) v$, would
exceed the actual physical velocity of the photosphere, $\dot{R}$, by
a factor,
\begin{equation}
{\Delta v\over \dot{R}} = \left({GM\over cR^2}\right)\left({P\over
2\pi}\right)=4.59 \left({M\over M_\odot}\right) \left({R\over
R_\odot}\right)^{-2}\left({P\over 1~{\rm yr}}\right) .
\label{relv}
\end{equation}
This implies that variable gravitational redshift must be considered
when interpreting the radial velocity noise introduced by photospheric
variability over long timescales in state-of-the-art searches for
low-mass planets \cite{Cegla}, such as the one described in
Ref. \cite{Mayor}. Interestingly, the apparent acceleration due to a
sinusoidal redshift variation is $90^\circ$ out of phase with the
apparent velocity and will appear like a planet on a circular orbit.

In some stars, the gravitational redshift signal could be counteracted
by a convective blueshift effect. Rising gas in convective cells is
hotter and brighter than sinking gas, causing a net blueshift in
absorption lines \cite{Dravinsf}. The resulting blueshift of $\sim
0.3~{\rm km~s^{-1}}$ for the Sun \cite{Dravins} and 0.2-$1~{\rm
km~s^{-1}}$ for other stars, could also impact spectroscopic planetary
searches \cite{Shporer}.

\paragraph*{Implications for stars in mini-halos.} The first stars in 
the Universe are thought to have formed due to molecular hydrogen
(H$_2$) cooling in mini-halos of dark matter with a velocity
dispersion of $\sim 1~{\rm km~s^{-1}}$ at redshifts $z\sim 10$--50
\cite{Loeb}.  Some of these early systems would have survived tidal
disruption if they reside in the outskirts of the Milky Way halo or
the Local Group.  Spectroscopic searches for dynamically-cold star
clusters with low velocity dispersions as tracers of the early
generation of star-forming mini-halos will have to correct for the
gravitational redshift effect of the constituent stars.

The measured extragalactic redshifts in cosmological surveys of
galaxies are expected to be systematically overestimated by the
gravitational redshifts of the emitting or absorbing components (stars
or gas clouds), supplemented by the redshift of their host system.
However, the redshift offset is typically negligible relative to the
measurement accuracy and only detectable in rich clusters of galaxies
\cite{Kaiser}, where the redshift induced by the cluster potential
dominates over that of the constituent stars. Interestingly, the
characteristic amplitude of potential fluctuations in the Universe,
$\vert \phi/c^2\vert \sim 10^{-5}$ is only a factor of $\sim 5$ larger
than the gravitational potential of the Sun $(GM_\odot/R_\odot
c^2)=2.12\times 10^{-6}$.

\paragraph*{Implications for other stellar systems.} Future astrometric
measurements of stars in the Milky Way galaxy by the forthcoming GAIA
mission \cite{GAIA} are expected to discover cold streams of stars
from the disruption of satellite galaxies in the Milky Way halo, as
well as wide binaries, and loose multiple star systems in its
disk. Follow-up spectroscopy can be used to detect the gravitational
redshift effect in these dynamically-cold systems, and potentially
open a new window for setting constraints on the mass-radius relation
of stars.

\bigskip
\bigskip
\paragraph*{Acknowledgments.}
I thank Eric Agol and Andy Gould for helpful comments on the
manuscript. This work was supported in part by NSF grant AST-1312034.

\end{document}